\documentclass[12pt]{article}
\usepackage[sumlimits,intlimits,namelimits]{amsmath}
\usepackage{amssymb}
\usepackage[english]{babel}
\usepackage{bbm}
\usepackage{parskip}
\global\parskip 6pt

\newcommand{\be}{\begin{eqnarray}}
\newcommand{\ee}{\end{eqnarray}}

\newcommand{\lb}{\label}
\def\>{\rangle}
\def\<{\langle}

\begin{document}

\begin{titlepage}
\hfill{LMU-ASC 53/11 }
\vspace*{.5cm}
\begin{center}
{\Large{{\bf Formation of Black Holes \\[.5ex]
in Topologically Massive Gravity}}} \\[.5ex]
\vspace{1cm}
Ivo Sachs
\\
{\em Center for the Fundamental Laws of Nature, Harvard University,\\ 
Cambridge, MA 02138, USA\\and\\Arnold Sommerfeld Center, 
Ludwig-Maximilians University,\\
Theresienstrasse 37, D-80333, Munich, 
Germany}\\
\end{center}
\vspace*{1cm}
\begin{abstract}
\noindent {We present an exact solution in 3-dimensional topologically massive gravity with negative cosmological constant which dynamically interpolates between a past horizon and a chiral AdS pp-wave. Similarly, upon time reversal, one obtains an AdS pp-wave with a future event horizon.}
\\
\vspace*{.25cm}за
\end{abstract}
\vspace*{.25cm}
\end{titlepage}
\section{Introduction}
The simplest theory of gravity allowing for genuine black hole solutions is  Einstein gravity in three dimensions with a negative cosmological constant. Although all classical solutions in this theory are locally equivalent to  three-dimensional anti-de Sitter space a so-called BTZ black hole is obtained as a discrete quotient \cite{BTZ} (see e.g. \cite{Carlip:1995qv,Birmingham:2001dt} for a review). The price paid for this simplicity is that there is no propagating graviton in this theory. This, however, can be remedied by adding a higher derivative term to the action in the form of a gravitational Chern-Simons term. This leads to so-called topologically massive gravity (TMG) \cite{Deser:1982vy}  which has a massive propagating degree of freedom (massive graviton) while retaining the BTZ black hole as a classical solution.  TMG has been the subject of considerable interest recently (see \cite{Witten:2007kt,Strominger,Carlip:2008jk,Grumiller:2008qz} for a partial list of references). 

Once we have a propagating graviton we are naturally lead to investigate the dynamics implied by this degree of freedom. Accordingly, an extensive discussion of quantum mechanical stability of TMG was initiated in \cite{Strominger,Grumiller:2008qz}. Similarly, linear classical stability of the BTZ black hole in TMG was established in \cite{Birmingham:2010gj} based on the mode analysis given in \cite{Sachs:2008gt}. In this note we extend the analysis in \cite{Birmingham:2010gj,Sachs:2008gt} by noting that some of the linear perturbations found in \cite{Sachs:2008gt} do in fact correspond to exact solutions of the full non-linear equations of motion of TMG. We then show that for a certain range of the topological mass parameter this solutions interpolate between a (singular) past horizon and a chiral pp-wave in anti-de Sitter space. 
The solution presented below then describes the full non-linear evolution of the linear perturbation of a white hole. Similarly, upon reversing the arrow of time, one obtains a chiral AdS pp-wave with  a (singular) future event horizon for $t\to\infty$. The idea that pp-waves could describe black holes has been pursued for some time (see e.g. \cite{Hubeny:2003ug} for a review) however, the result turned out to be negative in asymptotically flat space-time. In TMG with a negative cosmological constant this appears to be possible

\section{Non-linear Solution}
The action for topologically massive gravity with negative cosmological constant is given by
 \be\label{TMGA}
S&=&-\frac{1}{16\pi G}\int d^{3}x\sqrt{-g}\left(R+\frac{2}{l^2}\right)-\frac{1}{32\mu
\pi G}\int d^{3}x\sqrt{-g}\epsilon^{\lambda\mu\nu}\Gamma^{\rho}_{\lambda\sigma}
\left(\partial_{\mu}\Gamma^{\sigma}_{\nu\rho} +\frac{2}{3}\Gamma^{\sigma}_{\mu\tau}
\Gamma^{\tau}_{\nu\rho}\right),\lb{action}
\ee
where $\mu$ is the Chern-Simons coupling, and the parameter $l$ sets the scale of the cosmological constant of anti-de Sitter space, 
$\Lambda = -1/l^{2}$. 
The equations of motion then follow from 
\begin{equation}\label{varS}
\delta S = -\frac{1}{\kappa^2}\,\int d^3x\sqrt{-g}\,\delta g^{\mu\nu} \Big[G_{\mu\nu} + \frac1\mu\,C_{\mu\nu}\Big] =0.
\end{equation}
Here $G_{\mu\nu}$ is the modified Einstein tensor
\begin{equation}\label{ME}
G_{\mu\nu}=R_{\mu\nu}-\frac12\, g_{\mu\nu} R - \frac{1}{\ell^2}\, g_{\mu\nu}
\end{equation}
and $C_{\mu\nu}$ is the Cotton tensor
\begin{equation}\label{Cotton}
C_{\mu\nu}= \varepsilon_\mu{}^{\kappa\sigma}\nabla_\kappa \big(R_{\sigma\nu}-\frac14\, g_{\sigma\nu} R\big)\,.
\end{equation}
In the absence of the gravitational Chern-Simons term the only solution of the equations of motion is the space-time with constant negative curvature,  $AdS_3$. To continue we set $\ell=1$ so that  the Ricci scalar is given by  $R=-6$. Furthermore, taking a suitable quotient,  $AdS/\Gamma$, one obtains the topological BTZ black/white hole of mass $M=1$ with metric
\begin{equation}\label{btz}
ds^{2} = -\sinh^{2}\rho\; dt^{2} + \cosh^{2}\rho \;
d\phi^{2} + d\rho^{2},
\end{equation}
%
Clearly, this solution persists in the presence of the gravitational Chern-Simons term since both, the Einstein tensor and the Cotton tensor vanish separately.  However, there are more general solutions in this case (see e.g. \cite{Gibbons:2008vi,Chow:2009km,Chakhad:2009em} for a classification). Here we discuss a class of  solutions which in the far past approache a singular past horizon and asymptote to a pp-wave in the future. 

To see how this happens we first look at the linear perturbations of the BTZ background, that is, we write 
\begin{equation}\label{ans}
g_{\mu\nu}=\bar g_{\mu\nu}+h_{\mu\nu}
\end{equation}
As explained in \cite{Strominger} the non-trivial (not pure gauge) linear perturbations are given by the solutions of  
\begin{equation}\label{first}
\epsilon_\mu^{\;\;\alpha\beta}\nabla_\alpha h_{\beta\nu}+\mu
h_{\mu\nu}=0~~.
\end{equation}
For concreteness we take for $h_{\mu\nu}$
the left moving, ingoing mode (eqn 19 in \cite{Birmingham:2010gj})  at the future horizon of the BTZ black hole
\be\label{hl}
h_{\mu\nu}= e^{(1 +\mu)t }(\sinh \rho)^{1 +\mu} \left(\begin{array}{ccc}
1&1&\frac{2}{\sinh(2 \rho)}\\
1&1&\frac{2}{\sinh(2\rho)}\\
\frac{2}{\sinh(2\rho)}&\frac{2}{\sinh(2\rho)}&\frac{4}{\sinh^2(2\rho)} \end{array}\right).
\ee
Transforming to Finkelstein coordinates, $(u_{in/out}=t\pm\log \tanh(\frac{\rho}{2}),r=\cosh(\rho),\phi)$, it is not hard to see that (\ref{hl}) is regular at the past horizon for  $\mu> 1$ but satisfies the (generalized) Brown-Henneaux asymptotic boundary conditions \cite{Henneaux:2009pw} only for $\mu< 1$. Thus the BTZ black hole is linearly stable (see \cite{Birmingham:2010gj} for details). 

To proceed beyond the linear analysis we notice that the complete metric (\ref{ans}) nevertheless has a past horizon for $1>\mu>0$. More precisely, for 
\begin{eqnarray}\label{dis}
t=u_{out}+\log(\tanh(\frac{\rho}{2}))+ e^{(1 +\mu)u_{out}}f(\rho)\;,\qquad f(\rho)\propto \rho^{2\mu}(1+O(\rho^2))
\end{eqnarray}
one finds that (\ref{ans}) as well as its transverse curvature \cite{Barrabes:1991ng} and the geodesic velocities are continuous at the horizon while the component $R^{u_{out}}_r$ has a power-like divergence for $\mu <1$ and thus (\ref{ans}) approaches a singular null surface as $t\to -\infty$.

We now want to argue that (\ref{ans}) is, in fact, an exact  solution to the equations of motion (\ref{varS}). One way to see this is as follows: Let us compute the Ricci tensor for $g_{\mu\nu}$, that is 
\begin{equation}
R_{\mu\nu}[g]=\frac{R}{3}g_{\mu\nu}+\frac{\mu^2-1}{12}Rh_{\mu\nu}
\end{equation}
where $R=\bar R= -6$ in our conventions. The inverse metric is easily  found by noting that $h_{\mu\nu}=l_\mu l_\nu$, where $l_\mu$ is null with respect to $\bar g_{\mu\nu}$. Thus 
\begin{equation}
g^{\mu\nu}=[(\bar g+h)^{-1}]^{\mu\nu}=\bar g^{\mu\nu}-\bar h^{\mu\nu}
\end{equation}
 where $\bar h^{\mu\nu}=\bar g^{\mu\alpha}h_{\alpha\nu}$. 
 Thus
 \begin{equation}
R^\mu_{\;\;\;\nu}=\frac{R}{3}(\delta^\mu_{\;\;\;\nu}+\frac{\mu^2-1}{4}\bar h^{\mu}_{\;\;\;\nu}) 
\end{equation}
This exact expression is identical with that obtained in the linearized theory \cite{Sachs:2008gt}. Since $\bar h^{\mu}_{\;\;\;\nu}$ is traceless it then follows immediately that  $R=\bar R$. If we define 
\begin{equation}
S^\mu_{\;\;\;\nu}=R^\mu_{\;\;\;\nu}-\frac{R}{3}\delta^\mu_{\;\;\;\nu}=\frac{\mu^2-1}{4}\bar h^{\mu}_{\;\;\;\nu}
\end{equation}
we find that $S^\mu_{\;\;\;\nu}\propto l^\mu l_\nu$ and, in particular,  $S^2=0$. Thus $g_{\mu\nu}$ is of type $N$ in the Petrov classification (see e.g. \cite{Chow:2009km}). On the other hand $g_{\mu\nu}$ is of the Kundt-CSI type (see eg. \cite{Chakhad:2009em}). 
%

To see that this is an exact solution we first consider the modified Einstein tensor (\ref{ME}). We have  (with $R= -6$)
\begin{equation}
G_{\mu\nu}=\frac{1-\mu^2}{2}h_{\mu\nu}
\end{equation}
As for the Cotton tensor, only the first term in (\ref{Cotton}) contributes so that 
\begin{equation}
G_{\mu\nu}+\frac1\mu\,C_{\mu\nu} = \frac{1-\mu^2}{2}(h_{\mu\nu}+\frac{1}{\mu} \varepsilon_\mu{}^{\kappa\sigma}\nabla_\kappa h_{\sigma\nu})
\end{equation}
which, in turn, is recognized as the linearized equation for $h_{\mu\nu}$ in the background metric $\bar g_{\mu\nu}$. However, it can be shown that this equation is equally satisfied for the 'background' metric $\bar g_{\mu\nu}+h_{\mu\nu}$ thus confirming that (\ref{ans}) satisfies the exact equations of motion (\ref{varS}). 

\section{Interpretation}
Much of the mystery about this  non-linear solution is removed upon noting that $h_{\mu\nu}$ can be expressed in terms of a null Killing vector of $g_{BTZ}$, that is 
\be\label{hlk}
h_{\mu\nu}= f(t,\phi,\rho) \;\xi_\mu\xi_\nu
\ee
for some function $f$ and 
\be
\xi&=&e^{t+\phi}\left(-\coth(\rho)\partial_t+\tanh(\rho)\partial_\phi
+\partial_\rho\right)
\ee
is a null Killing vector\footnote{We should mention that $\xi$ not globally well defined in the BTZ background since it is not periodic in $\phi$.  In fact,  (\ref{ans}) does not admit a hyper surface orthogonal Killing vector anywhere and thus avoids the Birkhoff theorem \cite{Aliev:1996eh,Deser:2009er}. I would like to thank S. Deser for correspondence on this point.} of $g_{BTZ}$ and furthermore $L_\xi h=0$ \cite{Sachs:2008gt}.  We then choose coordinates   ($U,V,\lambda$)  such that $\xi=\partial_\lambda$, that is 
\begin{eqnarray}
t+\phi&=&-\log[\tanh(\rho)]+2U\nonumber\\
t-\phi&=&-\log [\sinh(2\rho)]+2V\\
\rho&=&\frac{1}{2}\hbox{arcosh}(e^{2V}\lambda)\nonumber
\end{eqnarray}
In these coordinates the BTZ metric with mass $M=1$ is given by 
\be
ds_{BTZ}^2=e^{2V}d\lambda dU+dV^2 +d U^2\,.
\ee
Our time-dependent solution (\ref{ans}) takes a similarly simple form 
\be\label{wgc}
ds^2=e^{2V}d\lambda dU+dV^2 +\left( 1+\frac{e^{(1+\mu)(U+V)}}{\sqrt{2}}\right)d U^2\,.
\ee
These coordinates are not particularly well suited to describe global properties of our solution since the coordinates are subject to identifications. However, they make it clear that our solution is just a special case of the class of solutions described in \cite{Deser:2004wd,AyonBeato:2004fq}. This class furthermore contains the AdS metric in Poincar\' e coordinates with $ds^2= e^{2V}d\lambda dU+dV^2$, as well as  the AdS plane-waves (see e.g. \cite{AyonBeato:2004fq,Carlip:2008eq}). We then see that our solution is a special case of an AdS-plane wave which asymptotes to a past horizon at early times. 

What about the mass and angular momentum of this solution? Given the fact that the metric (\ref{wgc}) falls off slower than the BTZ-background for large $V$ one might worry that the mass is ill-defined. However, it was shown in \cite{Henneaux:2011hv} that the surface integrals $Q[\partial_t]$  and  $Q[\partial_\phi]$ defining mass and angular momentum respectively are well defined and furthermore independent of the leading large $V$ contribution in (\ref{wgc}). In other words the mass is "time" independent. Thus  the singular null surface at $t\to-\infty$ evolves classically into a 
AdS-wave with mass $M=1$. In fact we can say more: Although $(\ref{ans})$ does not have a time-like killing vector we can define a surface gravity for the past  null surface w.r.t. to $\partial_t$ in the usual way. The so-defined surface gravity is constant on the null surface and furthermore agrees with that of  (\ref{btz}). Thus, while our pp-wave solution cannot be patched smoothly to the BTZ solution inside the horizon (see comments below (\ref{dis})) we can assign a temperature and  an entropy to the past null surface that agrees with that of an $M=1$ BTZ white hole. Furthermore, the geodesics can be extended continuously through the horizon.

So far we considered the classical evolution of a white hole but we can equally well describe the formation of a singular future horizon by considering the time-reversed situation. Due to the odd transformation property of the gravitational Chern-Simons term in (\ref{TMGA}) this requires that we change the sign of  $\mu$ at the same time. The corresponding solution which evolves from a AdS-wave in the past into a  future horizon for $t\to\infty$ is obtained form (\ref{hl}) upon substitution $t\to-t, \mu\to-\mu$. 

It would be desirable to better understand the causal properties of the interpolating non-linear solutions discussed here. This can be achieved by analyzing the geodesics on this time-dependent background. For instance one can show that, although the global charges of (\ref{wgc}) are the same as that for the non-rotating BTZ metric, there is an ergosphere \cite{Orsay}. We should also mention that the linear perturbation on which the exact solution is based is only one of an infinite tower of  solutions of the linearized equation. The generic solutions are "$SL(2,R)$ descendants" of (\ref{hl}). It would be interesting to see if these descendents do also give rise to exact solutions of the non-linear equations of motion. Finally, it would be interesting to consider the so-called logarithmic modes at $\mu=\pm 1$ \cite{Grumiller:2008qz}. We hope to report on these issues in a future publication. 
\\  \\

\noindent {\Large \bf  Acknowledgments}\\ \\
This research was supported in part by DARPA under Grant No.
HR0011-09-1-0015 and by the National Science Foundation under Grant
No. PHY05-51164,  by the Transregional Collaborative Research Centre TRR 33,
the DFG cluster of excellence ``Origin and Structure of the Universe'' as well as the DFG project Ma 2322/3-1.
I would like to thank A. Barvinski, S. Detournay, D. Grumiller, G. Ng, H. Reall, S. Ross, S. Solodukhin and W. Song for helpful discussions as well as KITP at UCSB for hospitality during the final stages of this project.

\end{document}